\documentclass{kapproc} % Computer Modern font calls

\usepackage{procps}

\usepackage[dvips]{graphicx}
\usepackage{wrapfig}
\usepackage{psfig}

\chapsectrunningheads

\upperandlowercase

 \let\footnote\savefootnote

\let\footnoterule\savefootnoterule

\normallatexbib %will produce bibliography entries as shown in the
\begin{document}

\articletitle {Is the Local Bubble dead?}

\chaptitlerunninghead{Panel Discussion: The Local Bubble} % Shorter running head title.

\author{Dieter Breitschwerdt,\altaffilmark{1} \& Donald P. Cox,\altaffilmark{2}}

\affil{\altaffilmark{1}Max-Planck-Institut f\"ur
Extraterrestrische Physik, Giessenbachstra{\ss}e, Postfach 1312,
85741 Garching, Germany\\
\altaffilmark{2} Department of Physics, University of
Wisconsin-Madison, 1150 University Ave., Madison, WI 53706, USA}

%% optional, to supply a shorter version of the title for the running head:
%%\chaptitlerunninghead{}

%\anxx{Beerends\, John G.}

\begin{abstract}
We give a summary of the current state of Local Bubble research,
resulting from the discussions of a dedicated panel meeting. After
more than 25 years of intense observational and theoretical work,
we are still far from a coherent picture, although a probable one
emerges at the horizon. A multi-supernova origin seems to be the
best guess, with non-equilibrium cooling and soft X-ray emission
accompanying its expansion. In addition our vantage point may
force us to accept a substantial but quantitatively unknown
contribution from heliospheric emission.
\end{abstract}


\section{Introduction}
%Here is the beginning of the article.\footnote{Here is a sample footnote.}
The Local Bubble Panel Session took place on the last day of the
Galactic Tertulia meeting, and therefore lots of new exciting
ideas that were spread during the meeting needed eagerly to be
discussed. One of them was a suggestion put forward by Rosine
Lallement, who together with her collaborator Barry Welsh, had
investigated the possible contribution to the soft X-ray emission
from charge exchange reactions of solar wind high ionization
stages with heliospheric plasma, a process that had been
successfully applied to X-ray emission of comets (Lisse et al.
1996; Cravens et al. 1997). We will discuss this in more detail
below. On the whole the panel discussion on the Local Bubble was
somewhat chaotic owing to active participation from the floor and
a belated sudden discovery that our departure from the conference
hall was soon required. This discussion was nominally led by
Dieter Breitschwerdt (who had prepared a list of topics with the
somewhat provocative title: ``Is the Local Bubble Dead?'') and Don
Cox (who played \emph{advocatus diaboli} whenever necessary), and
a number of interesting points were raised. Now with the leisure
of writing them up, perhaps we can make the significance more transparent.

One way of organizing the material is via the questions and puzzles
that remain, though perhaps for a broader audience, a survey of the
ideas currently afloat would be a better way to begin. Many of the
issues were discussed at length by Cox and Reynolds (1987). The
continuing development of Don's perspective is pretty well summarized
in three subsequent papers, one on the nature of the hot gas within the
Local Bubble (Smith and Cox, 2001), one on a possible origin of the
warm Local Fluff within the hot bubble (Cox and Helenius, 2003), and
one which discusses the possibility that the whole effort is a house of
cards (Cox, 2003). Dieter has written a few review papers
(Breitschwerdt 1996, 2001), emphasizing the problems of explaining EUV
and soft X-ray spectra by the conventional Local Hot Bubble model,
which is based on collisional ionizational equilibrium (CIE), and also
a paper on the possible origin of the bubble (Bergh\"ofer \&
Breitschwerdt 2002). Many others have contributed to the effort to
understand the local interstellar medium in general and the Local
Bubble in particular. Dieter organized a whole conference on the
subject in Garching in 1997. Participants at the Granada conference who
have wrestled with the subject include at least (we apologize for any
incompleteness): Avillez, Beckman, Breitschwerdt, Cox, Edelstein, Gry,
Hartquist, Helenius, Hurwitz, Korpela, Kuntz, Lallement,
Ma\'{i}z-Apell\'{a}niz, McCammon, Reynolds, Sanders, Shelton, Welsh.
Most of these are people who actually measure things, and have some
truth to tell. One person not at the Granada meeting, Priscilla Frisch,
has written extensively on the observational material available
concerning the local region. This should be a sufficient list from
which to start a literature search should one wish.

\section[The Situation in General Terms]
{The Situation in General Terms} \label{general}

\begin{wrapfigure}{r}[0pt]{61mm}
\centerline{
\includegraphics[width=0.9\hsize,bbllx=28pt,bblly=28pt,bburx=585pt,bbury=588pt,clip=]{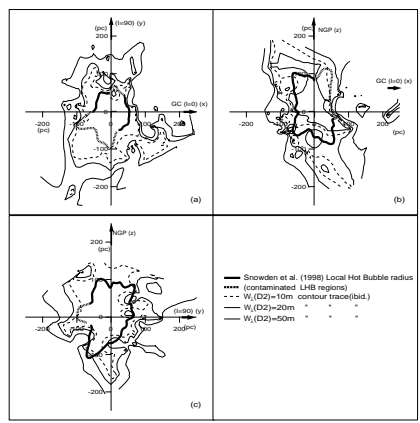}}
%\psfig{file=sfeir_99_fig6.ps,width=0.8\hsize,bbllx=30pt,bblly=30pt,bburx=581pt,bbury=585pt,clip=}}
% or use alternatively:
%    \epsfig{file=einstein_fig_01.eps,width=\hsize,bbllx=30,bblly=430,bburx=550,bbury=785,clip=}
% The bbllx,bblly,bburx,bbury ... are for the bounding boxes and should be used
% to delete any white space surrounding the figure
%\framebox[79mm]{\rule[-26mm]{0mm}{52mm}}
\caption{The local cavity as inferred from Na{\sc i} absorption line
studies by Sfeir et al.\ (1999) at different viewing angles. The
different contours mark different Na{\sc i} equivalent widths, with
$20$ m{\rm \AA} \, corresponding to $\log$(Na{\sc i}) = $11.0 \, {\rm
cm}^{-2}$ or $N_{\rm H} \approx 2 \times 10^{19} \, {\rm cm}^{-2}$. The
thick solid line represents the contour of the Local Bubble according
to ROSAT PSPC observations by Snowden et
al.\ (1998). %
 }
\label{nh}
\end{wrapfigure}

Soft X-rays reach the Earth in a pattern that shows a distinct
anti-correlation, in particular of 1/4 keV photons with the
distribution of local interstellar material (see Fig.~\ref{nh} for
Na{\sc i} absorption line studies and Rosine Lallement's updated figure
at this conference with a much larger number of stars). The emission is
diffuse, meaning that it does not arise from a collection of unresolved
point sources. The pattern is roughly what might be expected from
extragalactic emission, absorbed by galactic interstellar material,
except: A) it does not go to zero in the galactic midplane, and B) the
very lowest energy X-rays ought to be much more absorbed, but instead
show a very similar distribution. Therefore there must be a local
source of diffuse emission. Attempts to localize the source of the
emission have used shadowing by intervening material of putatively
known distance.  McCammon's thesis work showed that it arose closer
than the Magellanic Clouds.  Shadowing experiments done with ROSAT,
EUVE, and as reported by Dieter at this meeting with XMM-NEWTON, have
had more restrictive results. At high galactic latitude, there are
clouds of material several hundred parsecs from the Sun that shadow
\emph{part} of the X-ray emission. So, some arises within the first few
hundred parsecs (and much closer at lower latitudes where absorbing
and/or abutting walls of interstellar material are much closer) and at
high latitude, part arises from further away.%

This more distant emission is very patchy over the sky, and is the
origin of the idea that the Milky Way has a patchy distribution of hot
gas within its ``halo.''  The word halo is in quotes because Don
thinks of the disk of the Galaxy as reaching up a kiloparsec or
more and that this emission is likely found within it, not in some
vast region beyond.  This particular point is open to dispute, but
will likely not be settled until we have a better understanding of
why the very lowest energy emission does not have a substantially
different distribution on the sky. It was early on realized that
this emission resembled that from a hot interstellar plasma with a
temperature of roughly $10^6$ K. Plasma emission models based on
CIE suggested that the surface brightness could be achieved in a
volume extending 100 pc with a thermal pressure comparable to that
expected in a large hot cavity in equilibrium with its
surroundings, $p/k \sim 10,000-20,000 \, {\rm cm}^{-3}\, {\rm K}$. The
so-called ``displacement'' or Local Hot Bubble (LHB)'' model, put
forward almost simultaneously by Wisconsin (Sanders et al.
1977) and a Japanese group (Tanaka and Bleeker 1977) could explain
satisfactorily  at that time the local soft X-ray emission by
claiming that \emph{all} of the emission in the 1/4 keV band was
due to a hot plasma of $\sim 10^6$ K and $n_e \sim 5 \times
10^{-3} \, {\rm cm}^{-3}$, displacing neutral surrounding
gas\footnote{When the first ROSAT shadow of the Draco Nebula was
discovered (Snowden et al. 1991, Burrows \& Mendenhall 1991) it
became evident that roughly 50\% of the 1/4 keV emission was
arising from beyond the cloud, i.e. at a minimum distance of
300~pc and thus far beyond the Local Bubble.}. A roughly
contemporaneous result was that absorption line studies were
finding that the region around the Sun in fact has very little
interstellar material for the nearest hundred parsecs or so,
depending on direction.  (As mentioned above, the papers by Lallement and by Welsh at
this meeting show the current state of mapping that cavity.) The cavity has a very
irregular geometry, but extends to several tens of parsecs in the
galactic midplane and opens at high (and low) galactic latitudes
where it extends further than has been mapped so far, several
hundred parsecs (see Fig.~\ref{nh}). And, it is precisely in those
seemingly open directions that EUVE was able to detect
extragalactic objects.

\section[``The Devil is in the Details'']{``The Devil is in the Details''}
\label{devil}
Because of its extreme faintness and the technological limitations on
spectral resolution, the soft X-ray background has been slow to reveal
its full spectral character.  The measurements are often compared with
those expected from hot plasmas with solar abundances, by calculating
CIE emission spectra (vs. T and absorbing column) and folding the
latter through the instrumental response.  Best fits are made in the
usual way, with one or two emitting temperature components. The spectra
that produced the best fits to the 1/4 keV X-rays also predicted softer
emission, particularly a very bright iron complex around 72 eV. They
also predicted little oxygen K-shell emission.  Several attempts to
verify the brightness of the soft iron lines failed to detect them,
until finally they seem to have been seen by XQC at roughly 1/7th of
the initially anticipated brightness--see paper by McCammon in this
volume. To continue spectral fitting in the above way, one must assume
that iron, at least, and possibly other elements remain somewhat
depleted in their abundances in the hot gas. (Smith and Cox, 2001,
showed that this is feasible so long as the gas has not been heated too
many times.)  The spectral results of DXS, on the other hand cannot be
fit very well by this process (Sanders et al. 1998), seeming to require
considerable improvement in the plasma modeling codes or, perhaps an
alternative source type for a significant fraction of the
emission--more below on this point.

At the higher energies of the 3/4 keV band, Snowden et al. (1993)
analyzed ROSAT PSPC data towards MBM12, commonly believed to be a
medium latitude cloud within the Local Bubble at a distance $d$ of
$58\pm5 < d <  90\pm 12$ pc (Hearty et al.\ 2000; but see recent
photometric analysis of M dwarfs by Luhman (2001) who place it out to
275 pc just to add more devilish details to the puzzle). They found
that all of the 3/4 keV band emission seemed to come from beyond the
cloud, and set a fairly strong upper limit to the foreground emission,
that attributed to the hot Local Bubble, and to its temperature (found
as described above).  On the other hand, together with Michael
Freyberg, Dieter reanalyzed these data, as well as data from the Aquila
molecular cloud, which is one of the darkest nearby regions of the
diffuse X-ray sky, and located in almost the opposite direction. On the
basis of these data, including higher energies than analyzed by Snowden
et al., they derived a new value for a local component which is $kT =
0.18$ keV ($\sim 2.1 \times 10^6$ K), much higher than previous values.
There was clearly cause for some disagreement.

\begin{figure}[htbp]
%
%\vspace{-5mm}
\begin{center}
\begin{minipage}{5.0cm}
\vspace*{0mm}
\includegraphics[width=\textwidth,clip]{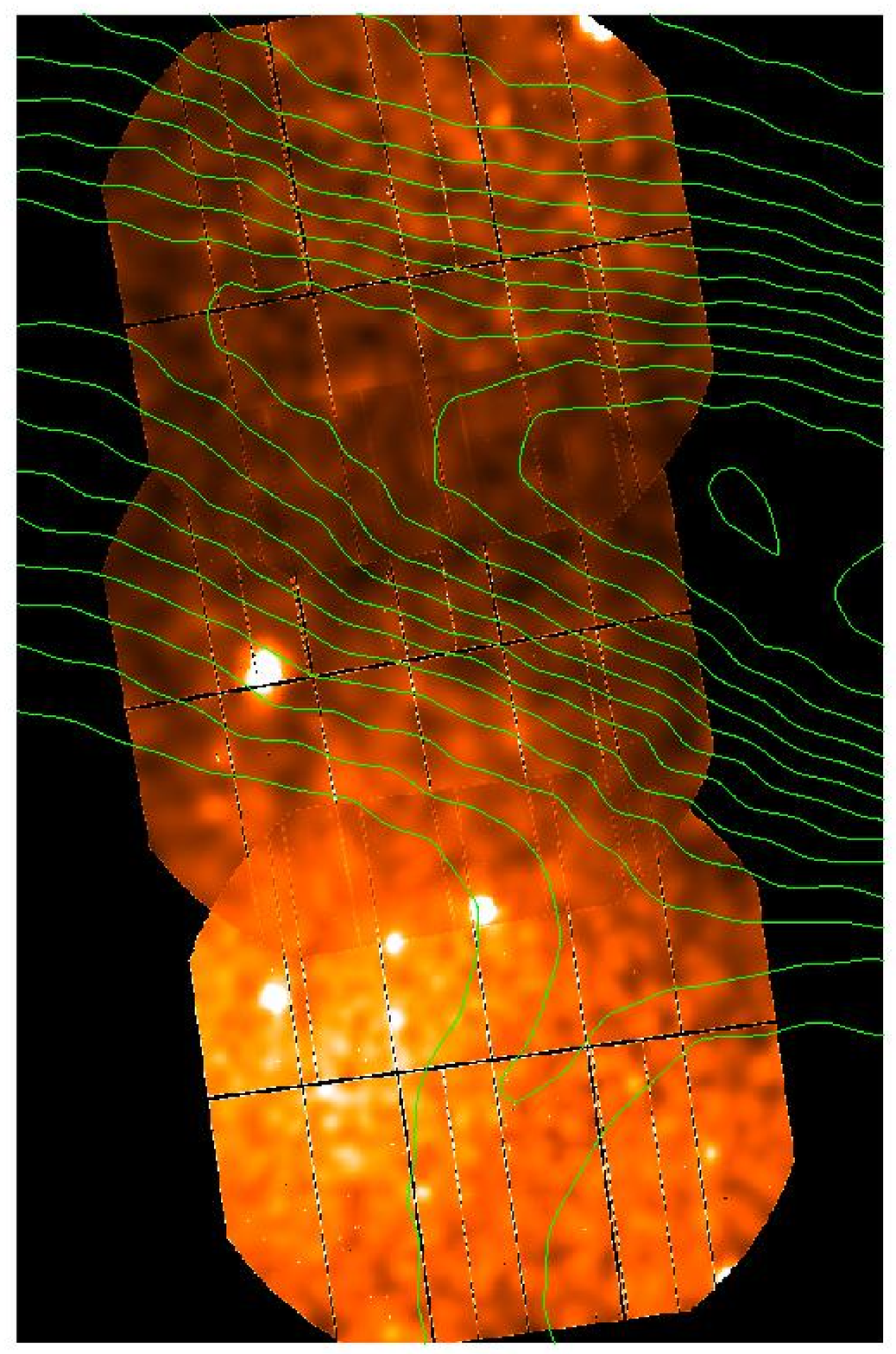}
\caption{Mosaic of three individual pointings of the Ophiuchus
molecular cloud, showing the first XMM-Newton X-ray shadow (in the
range $0.5 - 0.9$ keV). There is a clear anticorrelation between soft
X-ray emission and the overlaid IRAS 100$\,\mu$m contours. The color
coding represents the X-ray intensity with white being the maximum.}
%\vspace*{1mm}
\end{minipage}
\hspace*{1cm}
\label{oph-ima}
\begin{minipage}{5.5cm}
\vspace*{1mm}
\includegraphics[width=\textwidth,bb= 23 26 184
148,clip]{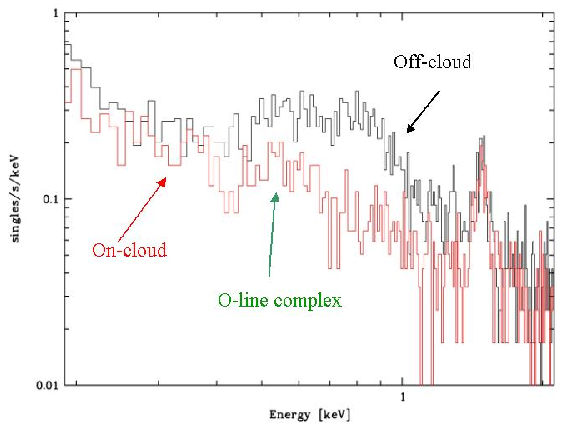}
\caption{Spectra (in counts/s/keV) towards the Ophiuchus cloud as
derived from two 20 ksec XMM-Newton EPIC pn observations. Emission line
complexes are clearly distinguishable at $0.5 - 0.7$, and $\sim 0.9$
keV, and to a minor extent at $\sim 0.3$ keV. The on-cloud pointing
(lower curve) contains mainly emission from the Local Bubble, while the
off-cloud (upper curve) observation has also significant contributions
from the Loop I superbubble showing up as $0.8 - 0.9$ keV emission
(Fe-L complex), and arising from higher temperatures there.
 }
\end{minipage}
\label{oph-sp}
\end{center}
\end{figure}

\begin{wrapfigure}{R}[0pt]{70mm}
%\begin{minipage}[t]{80mm}
 \centerline{\includegraphics[width=\hsize,bb= 20 18 460 368,clip]{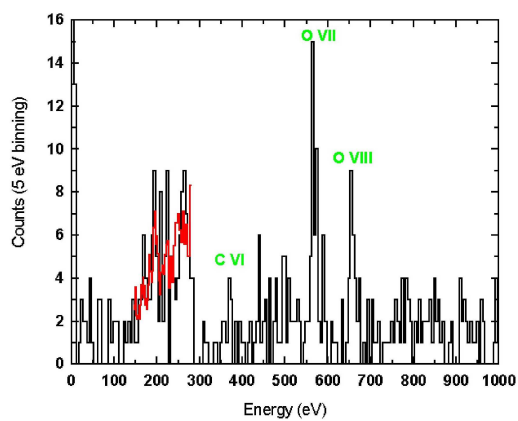}}
\caption{High resolution spectrum of the diffuse soft X-ray
background (black histogram, arbitrary units) in the 70 - 1000 eV
energy range, taken with the XQC spectrometer during a 100\,s
sounding rocket flight (McCammon et al. 2002). The instrument FOV
was about 1 sr and centred at $l=90^\circ$, $b=60^\circ$.
Prominent emission lines are marked. The DXS spectrum taken from a
region near the Galactic plane (Sanders et al. 2001) is superposed
(grey or red lines, respectively, in the energy range $150 - 284$ eV) for comparison. %
} \label{xqc_spec}
%\end{minipage}
%
\end{wrapfigure}

The latest XMM-Newton observations done by the group in Garching (MPE)
show the presence of emission near 0.56 keV and 0.65 keV (indicating
O{\sc vii} and O{\sc viii}, respectively) for all targets, situated in
different directions and at different distances and latitudes in the
sky, e.g.\ MBM12, G133-69 (in the southern hemisphere), North Galactic
Pole Rift (NGP) and the Ophiuchus molecular cloud (see
Figs.~\ref{oph-ima} and \ref{oph-sp}). This is fully consistent with
the X-ray quantum calorimeter (XQC) results (see Fig.~\ref{xqc_spec})
by McCammon et al. (2002). Although the spectral resolution of XQC is
much higher than of the EPIC pn camera, the short duration of the
sounding rocket flight required observing a region that covered both
Local Bubble and Galactic halo. The shadowing observed, for example in
Fig.~\ref{oph-ima} allowed separation of the local and distant
emission, and a two component spectral fit to the foreground yields
temperatures of 0.08 and 0.14 keV, respectively.

The bottom line at present appears to be that the spectrum has less
emission from the iron complex around 72 eV, other characteristics
suggestive of maybe some depletion of refractory elements, and somewhat
more oxygen K-shell emission than expected from a contemporary single
temperature solar abundance plasma model assuming CIE that fits best
the softer emission.

Robin Shelton reminded us of more problems lurking in the back by
mentioning recent FUSE ($905 < \lambda < 1195 $\AA) data: inside
the Local Bubble the 2$\sigma$ upper limit on the surface brightness
of O{\sc vi} (resonance line emission at $\lambda\lambda 1032,
1038$\AA) is extremely low, at most $530$ and $500 \, {\rm ph}
\,{\rm cm}^{-2} \, {\rm s}^{-1} \, {\rm sr}^{-1}$, respectively
(Shelton 2003), in disagreement with all current models. She also
thinks that such a low limit would rule out the possibility, put
forward by John Raymond at this conference, to explain the
complicated spectra by a superposition of different X-ray emitting
and absorbing chunks of gas within the line of sight.

\section[Whence the Local Bubble]{Whence the Local Bubble}
\label{lb_origin}
Determining the spectrum of the soft X-ray background is just a first
step in trying to understand the origin and the further evolution of
the bubble. Models are needed too.  The energy content of the hot gas
bubble (The Local Bubble, as distinct from The Local Cavity) described
before is roughly that of one supernova, and various models for the
reheating of a portion of the Local Cavity by a relatively recent
supernova have been made. The earliest models, as the one by Cox \&
Anderson (1982), assumed that the remnant was young, about $10^5$ years
old, so that its expansion velocity would give a post shock temperature
of $10^6$ K. Those models were found to have various problems and
gradually gave way to attempts to model the bubble as more like a
``Slavin Bubble'', the old hot remnant of a supernova explosion (cf.
also Innes \& Hartquist, 1984) that has largely equilibrated in
pressure with its surroundings. The Smith and Cox (2001) paper referred
to above is a recent version. The reheating supernova was inferred to
have occurred roughly 3 Myr ago, somewhere in the vicinity of the Sun.
Independent evidence for the occurrence of such a supernova (cf. Knie
et al.\footnote{These authors have analyzed the ferromanganese crust of
deep ocean layers, and found an enhancement of $^{60}$Fe consistent
with an explosion at about 5 Myr ago.} 1999) was cited with glee.
Until recently, the spectral confirmation that the X-ray emission
actually arises from hot gas has been very poor. Some expected spectral
lines have now been seen, but not all characteristics of the observed
emission are consistent with existing models of hot gas. Encouraged by
some problems in the interpretation of
data (e.g.\ the region of ''bizarre emptiness'' as Don put it
towards $\beta$CMa, or the thermal pressure imbalance between the LHB and
the Local Fluff, a partially ionized cloudlet surrounding the
solar system, as well as a probable inconsistency between EUV and
soft X-ray data) Breitschwerdt and Schmutzler (1994) seized upon
this weakness to propose an alternative model that
bedevils\footnote{These are Don's words; Dieter agrees that the
model is somewhat bold, owing to his youth at that time, but still
believes that the Local Bubble spectrum has to be some kind of
``non-equilibrium'', even if it were as extreme as produced by
Solar Wind charge exchange reactions in the heliosphere.} us to
this day. In this model, relatively high density material heated
by one or more supernovae expanded very rapidly into a surrounding
low density region, cooling adiabatically to low temperature as it
did so. The expansion was so rapid that high stages of ionization
characteristic of the initially high temperature were ``frozen
in'' as occurs in the Solar Wind. They proposed that the
subsequent recombination of those high ions to be the source of
the soft X-ray background. A telling feature of such a
recombination spectrum was expected to be the recombination
continua, which at such low temperatures would resemble asymmetric
emission lines at the recombination edges.

One can get into quibbles about the likelihoods of various
scenarios, but it is safer to examine the predictions and the data
and to try to understand what we are seeing. It is likely that the
Breitschwerdt and Schmutzler scenario has occurred somewhere. To
decide whether it is related to our soft X-ray background,
however, we need good spectra. The current status of obtaining
such spectra was summarized at this meeting by McCammon or Sanders, or both. We
have DXS and XQC, and we desperately need a SMEX or MIDEX. Nobody
is just going to hand us the data, even though detector
development has reached the point that the crucial questions can
be addressed.

Meanwhile, the whole field has been complicated by the strong
possibility that a Breitschwerdt-like mechanism is operating to create
much of the soft X-ray background within the Solar System! Those high
stages of ionization in the Solar Wind mentioned earlier can undergo
charge exchange on neutral atoms, and the subsequent cascade creates
copious X-rays. This mechanism is believed to be responsible for the
extremely bright X-ray emission from comets in the inner Solar System.
Comets do not supply the only neutral targets for charge exchange in
the Solar System, however. There are now strong indications that a
temporally fluctuating component that was spectrally indistinguishable
by ROSAT from the soft X-ray background (the ``Long Term
Enhancements'') derives from Solar Wind charge exchange on the
exosphere of the Earth! But the Earth is not the only other target.
Using estimates of the charge exchange cross sections, it is easy to
calculate the approximate level of X-ray emission from charge exchange
of Solar Wind ions encountering interplanetary neutrals (e.g. Cox,
1998). The answer is frighteningly large! The current situation is
presented in Lallement's paper in this volume, where she has integrated
the solar wind flux against the interstellar distributions of hydrogen
and helium, \emph{in the directions and for the times at which ROSAT
was making the survey}. This provides a spatial template against which
to measure the potential contamination. Her paper goes on to assume
various normalizations, based on the poorly known cross sections and
ionic populations, and then subtracts off the heliospheric contribution
to see what is left of the Local Bubble. She points out that what is
left correlates better with the structure of the spatial cavity
(discussed above) than did the total SXRB.

How much is she subtracting? Within the uncertainties, \emph{all}
of the SXRB in the galactic midplane could be heliospheric! This
still leaves quite a lot at high (positive and negative) latitudes
that can't be heliospheric, along with some extended regions at
lower latitudes that correlate with the surrounding structure. She
seems to prefer to preserve the Local Bubble, perhaps to preserve
those lobes as part of it, and therefore to propose subtracting
only, say 50 to 80\% of the emission in the lowest emission
directions. This is as good and responsible a guess as anyone can
make at this time.  We need both better heliospheric emission
models and measured spectra before much progress can be made.

Does this require a disastrous change in our view(s) of the hot
Local Bubble? She argues not. Suppose, for example, that in some
low latitude directions the Local Bubble emission is only 25\% of
what we previously thought. As emissivity goes as density squared,
it would imply that the density and pressure were halved. At high
latitudes, where the subtracted fraction is much less, a four
times greater path length would be required to get the observed
brightness, and the extension of the Local Bubble to beyond the
shadowing clouds would be a fairly natural consequence. The total
energy in hot gas may then exceed that which one supernova could
supply, but living in a very large bubble makes multiple
supernovae much more likely.

Dieter has ventured to construct such a model (Bergh\"ofer \&
Breitschwerdt 2002), and interestingly, at around the same time a
similar idea was put forward by Ma\'{i}z-Apell\'{a}niz (2001), who also
presented his views during the panel session. The basics are very
simple. As has been mentioned above, the X-ray emissivity of the Local
Bubble can easily be accounted for by one or two recent supernovae,
whereas it took considerably more explosions to blast the Local Cavity
free of gas; about 10 - 20 supernovae can do this. But there are no
early type stars within the Local Bubble. On the other hand there are
plenty in the nearby Sco-Cen association. Stellar kinematics data show
that one or several moving groups of young stars have passed through
the Local Bubble on their way to Sco-Cen. Fitting an IMF appropriate
for galactic OB associations to such a group, Bergh\"ofer \&
Breitschwerdt (2002) were able to calculate the sites of explosions and
the intervals between them. Recent high resolution simulations on a
large grid (including the galactic fountain, for details see the
contribution of M. Avillez, this volume) and occurring in an
inhomogeneous background, disturbed by previous generations of
supernovae have shown (Avillez \& Breit\-schwerdt 2003) that both the
morphology and the timescales of the Local Bubble and the Loop~I
superbubbles are consistent with observations.

The numerical simulation presented of this activity was quite
striking, and at several times appeared to show filaments or
sheets of material intruding into the hot gas. Sometimes these
appeared to be of the sort advocated by Frisch, material ejected
from near the wall by recent explosion activity there. At other
times, they may have arisen from shear flows in the highly
convective medium. What could not be discerned in the time
available was whether they occasionally derived from a mechanism
proposed recently by Cox and Helenius (2003), filaments of
material pulled from the boundary through the hot gas by magnetic
tension. At least one of the panelists thought this had to be the
case. Distorted magnetic fields of the magnitude observed in the
interstellar medium are very quick to straighten themselves out
when they have very little material on them, with only
hydrodynamic drag through the surroundings to slow them down. This
view was discussed briefly at one of the sessions and argued
against on the basis of the resulting ionization of helium and
argon, compared to the observations. This led to discussion of the
sorry state of our knowledge of low temperature dielectronic
recombination in particular, and recombination of things like
argon in general. Enough smoke, and the issue was left hanging.

This question of the ionization in the Local Fluff fascinates
a number of people, as does the issue of the
apparent thermal pressure imbalance between the Local Fluff
and the surrounding hot gas.  Attempts to cure the
latter with a strong magnetic field in the Fluff run into
constraints from Voyager 1 not yet having encountered
the Solar Wind termination shock.  Having run out of space,
we refer the reader to discussions of these matters
in Cox and Helenius (2003).  (Dieter wants to point out that
neither of these problems exists in his bedeviling
model, however.)

\section[Conclusions]{Conclusions}
\label{conc}
The Local Bubble is still an active and highly relevant research
topic and will keep surprises for the people who venture to learn
more about it. From time to time there will be bold attempts to
get rid of it, as it has happened in the past, when it was thought
to be an interarm region, or now, that some believe that much of
its X-ray emission could be of heliospheric origin. Yet the cavity is
still with us. We have both been working long enough in the field
to think that the emblem for the City of Paris will also hold for
the future of the Local Bubble: \emph{fluctuat nec mergitur!}

\begin{acknowledgments}
Don Cox acknowledges support of this work by NASA's Astrophysics Theory
Program under grant NAG5-12128. Dieter Breitschwerdt acknowledges
support from the Bundesministerium f\"ur Bildung und Forschung (BMBF)
by the Deutsches Zentrum f\"ur Luft- und Raumfahrt (DLR) under grant 50
OR 0207 and the Max-Planck-Gesellschaft (MPG). We would like to thank
Emilio Alfaro, Pepe Franco and the SOC and LOC for organizing an
exciting conference.
\end{acknowledgments}


\begin{chapthebibliography}{}
\bibitem{} Avillez, M.A., Breitschwerdt, D. 2003, RevMexAA 15, 299
\bibitem{} Bergh\"ofer, T.W., Breitschwerdt, D. 2002, A\&A 390, 299
\bibitem{} Breitschwerdt, D. 1996, Space Sci.\ Rev. 78, 173
\bibitem{} Breitschwerdt D., 2001, ApSS 276, 163
\bibitem{} Breitschwerdt D., Schmutzler T. 1994, Nature 371, 774
\bibitem{} Burrows, D.N., Mendenhall, J.A. 1991, Nature 351, 629
\bibitem{} Cox, D. P. 1998, in IAU Colloq. 166, {\it The Local Bubble and Beyond},
eds.\ D. Breitschwerdt, M. J. Freyberg, \& J. Tr\"umper Lecture
Notes in Physics (Berlin: Springer), 506, 121
\bibitem{} Cox, D.P. 2003, in {\it From Observations to Self-Consistent Modelling
of the ISM in Galaxies}, eds. M. Avillez \& D. Breitschwerdt,
Kluwer, ApSS (in press), astro-ph/0302470
\bibitem{} Cox D.P., Anderson, P.R. 1982, ApJ 253, 268
\bibitem{} Cox D.P., Helenius, L. 2003, ApJ 583, 205
\bibitem{} Cox, D. P., \& Reynolds, R. J. 1987, ARA\&A, 25, 303
\bibitem{} Cravens, T.E. 1997, Geophys.\ Res.\ Lett.\ 24, 105
\bibitem{} Hearty, T., Fern\'andez, M., Alcal\'a, J.M.,
Covino, E., Neuh\"auser, R. 2000, A\&A 357, 681
\bibitem{} Innes, D.E., Hartquist, T.W. 1984, MNRAS 209, 7
\bibitem{} Jelinsky, P., Vallerga, J.V., Edelstein, J. 1995, ApJ 442, 653
\bibitem{} Knie, K., Korschinek, G., Faestermann, T., et al. 1999,
Phys.\ Rev.\ Lett. 83, 1
\bibitem{} Lisse, K.C., et al. 1996, Science 274, 205
\bibitem{} Luhman, K.L. 2001, ApJ 560, 287
%\bibitem{} Lyu, C.H. \& Bruhweiler, F.C. 1996, ApJ, 459, 216
\bibitem{} Ma\'{i}z-Apell\'{a}niz, J. 2001, ApJ 560, L83
\bibitem{} McCammon, D., Almy, R., Apodaca, E., Bergmann-Tiest, W., et al., 2002, ApJ 576, 188
\bibitem{} Reynolds, R. J., \& Cox, D. P. 1992, ApJ, 400, 33
%\bibitem{} Press, W. H., et al. 1989, Numerical Recipes: The Art of Scientific Computing (New York: Cambridge Univ. Press)
\bibitem{} Sanders, W.T., Edgar, R.J., Kraushaar, W.L., McCammon, D., Morgenthaler, J.P., 2001,
ApJ 554, 694
\bibitem{} Sanders, W.T., Edgar, R.J., Liedahl, D.A., Morgenthaler, J.P.
1998, in IAU Colloq. 166, {\it The Local Bubble and Beyond}, eds.\
D. Breitschwerdt, M. J. Freyberg, \& J. Tr\"umper Lecture Notes in
Physics (Berlin: Springer), 506, 83
\bibitem{} Sanders, W.T., Kraushaar, W.L., Nousek, J.A., Fried, P.M. 1977,
ApJ 217, L87
\bibitem{} Sfeir, D. M., Lallement, R., Crifo, F.,\& Welsh, B. Y. 1999,
A\&A, 346, 785
\bibitem{} Shelton, R.L. 2003, ApJ 589, 261
\bibitem{} Slavin, J.D., Frisch, P.C. 1998, in IAU Colloq. 166, {\it The Local Bubble and Beyond}, eds.\ D.
Breitschwerdt, M. J. Freyberg, \& J. Tr\"umper, Lecture Notes in
Physics (Berlin: Springer), 506, 305
\bibitem{} Smith, R.K., Cox, D.P. 2001, ApJS 134, 283
\bibitem{} Snowden, S.L., Egger, R., Finkbeiner, D.P., McCammon, D.,
et al., 1998, ApJ 493, 715
\bibitem{} Snowden, S.L., McCammon, D., Verter, F., 1993, ApJ 409, L21
\bibitem{} Snowden, S.L., Mebold, U., Hirth, W., Herbstmeier, U.,
et al. 1991, Science 252, 1529
\bibitem{} Tanaka, Y., Bleeker, J.A.M. 1977, Space Sci.\ Rev.\ 20, 815
\bibitem{} Walters, M. A. 1993, B.S. thesis,
University of Wisconsin, Madison
\bibitem{} Welsh, B.Y., Sfeir, D. M., Sirk, M. M.,\& Lallement, R. 1999,
A\&A, 352, 308

\end{chapthebibliography}
\end{document}